# Synthesis and magnetic properties of a quantum magnet PbCuTeO$_5$

B. Koteswararao,[1] S. P. Chilakalapudi,[1] Aga Shahee,[2] P. V. Srinivasarao,[3] S. Srinath,[1] A. V. Mahajan[2]

[1]*School of Physics, University of Hyderabad, Central University PO, Hyderabad 500046, India.*
[2]*Department of Physics, Indian Institute of Technology Bombay, Mumbai 400076, India.*
[3]*Department of Physics, Acharya Nagarjuna University, Nagarjuna Nagar 522 510, India.*

**Abstract**

We report the structural and magnetic properties of a quantum magnet PbCuTeO$_5$. The triclinic structure of PbCuTeO$_5$ comprises of alternating layers (*ab*-planes), in which one layer is composed of $S = ½$ dimer chains and another layer is composed of $S = ½$ trimer chains formed by corner-shared CuO$_4$ units. Magnetic susceptibility $\chi(T)$ data show a Curie-Weiss behavior with an antiferromagnetic Curie-Weiss temperature $\theta_{CW}$ of -165 K. At low temperature, both the heat capacity $C_p(T)$ and $\chi(T)$ show an anomaly at $T_c \approx 6$ K with a weak ferromagnetic (WFM) moment suggesting the appearance of long-range order. The magnetization *vs.* field $M(H)$ data at 2 K also provide evidence for WFM behavior. Magnetic frustration with a frustration parameter $f = \theta_{CW}/T_c$ of about 27 is observed. Magnetic specific heat data suggest the presence of a large entropy in the paramagnetic region, well above $T_c$, suggesting the presence of short-range spin correlations. The observed results might originate from the frustrated network of $S = ½$ distorted checkerboard lattice formed due to the coupling of the spin chains *via* TeO$_6$ octahedral units in the *ab*-plane.

## I. INTRODUCTION

Geometrical frustration can forbid the formation of magnetic long-range order (LRO) even at $T = 0$ K and hence lead to exotic spin ground states such as quantum spin-liquids, according to the initial proposal by P. W. Anderson [1]. The experimental realization of new spin-liquid materials is currently one of the central topics in condensed matter physics. The spin-liquid behavior can be realized in two-dimensional (2D) geometrically frustrated magnetic systems (GFMS) including triangular (edge-shared), Kagome (corner-shared), and checkerboard lattice models [2-4]. In the recent past, Herberthsmithite ZnCu$_3$(OH)$_6$Cl$_2$, a celebrated Heisenberg antiferromagnet pertaining to the $S = ½$, 2D Kagome model, has been the focus of attention [5-10]. In this material, the Cu$^{2+}$ ions ($S = ½$) lie on the vertices of a perfect corner-shared triangular lattice, *i.e.*, the Kagome lattice. Magnetic susceptibility data analysis yielded the antiferromagnetic (AFM) Curie-Weiss temperature ($\theta_{CW}$) of about -300 K and the exchange coupling of about $J/k_B \approx -180$ K. No signature of magnetic LRO was found down to 50 mK, claiming that it is a potential candidate for spin-liquid. On the other hand, the organic salt K-(BEDT-TTF)$_2$Cu$_2$(CN)$_3$ was shown to exhibit a gapless spin-liquid state firmly establishing its candidature to the class of 2D edge-shared triangular lattice [11, 12]. While the study of the spin-liquid state in the triangular and Kagome model materials has been extensively advancing, no materials have yet been reported in the context of checkerboard and/or 2D pyrochlore lattice model which has been predicted theoretically as good candidates for realizing the spin-liquid, valence bond crystal, *etc.* [13-17].

In this paper, we introduce a new quantum spin system PbCuTeO$_5$. According to its Cu-O-Cu exchange paths, this compound appears to have spin chains lying in its crystallographic *ab*-plane. However, magnetic data do not exhibit any characteristic feature of spin chain magnetism. Magnetic susceptibility data rather follow a Curie-Weiss behavior with antiferromagnetic Curie-Weiss temperature ($\theta_{CW}$) of about -165 K. At low-*T*, the system undergoes a transition at $T_c$ = 6 K. Specific heat data also show the evidence of large magnetic entropy above $T_c$. The observed results suggest that frustrated magnetism in the titled system could be originated from a 2D checkerboard lattice (which is one of the model system for highly frustrated magnetic lattice) with anisotropic couplings in the *ab*-plane, which forms due to the coupling between spin chains *via* TeO$_6$ units in the *ab*-plane.

## II. EXPERIMENTAL DETAILS

The polycrystalline PbCuTeO$_5$ sample was prepared by the conventional solid-state reaction method. The molar mixtures of PbO, CuO, and Te with a ratio of 1:1:1.05 were ground together, pressed into pellets. The additional 5% of Te was added to compensate for the loss due to evaporation during synthesis. We have then fired the sample in air successively at temperatures of 500°C, 600°C, and 720°C. Each firing was done for 12 hrs with an intermediate grinding. Finally, the firing was carried out at 720°C for 24 hrs to get the final phase of the compound. The powder x-ray diffraction measurements were done at room temperature using a Rigaku diffractometer equipped with the monochromator for Cu-$K_{\alpha 1}$ radiation. The magnetic measurements were performed in the temperature *(T)* range of 2 to 300 K and in the magnetic field (*H*) using a SQUID-VSM from Quantum Design. Specific heat measurements were performed by 2τ relaxation method using a Physical Property Measurement System (PPMS) from Quantum Design.

## III. RESULTS AND DISCUSSION

### A) X-ray diffraction and structural features

To check for the formation of single phase of PbCuTeO$_5$ sample, the measured powder XRD data was refined with FullProf Suite program [18] using the initial parameters given by Pavlina Choleva [19]. As shown in Figure 1, the refinement of XRD indicates a good fit with the $\chi^2$ value of 6.95. The lattice parameters of this triclinic unit cell are obtained to be *a* = 6.431(4) Å, *b* = 11.317(7) Å, *c* = 12.31(9) Å, α = 107.9°, β = 90.9°, and γ = 90.4°, which are in agreement with those of previously reported values [19]. The obtained atomic positions are mentioned in the Table 1.

The PbCuTeO$_5$ compound has triclinic unit cell with the space group *P*-1 (No. 2). As shown in Figure 2, the structure is built by TeO$_6$ octahedra, CuO$_4$ distorted squares, and Pb atoms. The Cu atoms lie only in the *ab*-planes which are in fact well separated from each other by an inter-planar distance of 6.15 Å, thereby forming 2D planes (as shown in Figure 3). The structure has two kind of layers; one at *z* = 0 and the other at *z* = 0.5. The *z* = 0 layer is built by dimer chains, while the *z* = 0.5 layer is built by trimer chains, as shown in Figure 3. The bond angles and bond lengths between Cu-Cu atoms are mentioned in Table II. According to the super-exchange (SE) pathways between Cu atoms *via* O atoms, the bond angles are in the range of 110º to 120º degrees, which likely favors antiferromagnetic bonds according to the Goodenough rules [20]. With the consideration of magnetic super-super-exchange (SSE) bonds between Cu atoms *via* O–Te-O, it is to be suggested that these layers are forming like a distorted checkerboard lattice. As it can be seen from the Figure 3 that there are two types of checkerboard lattices formed with dimer and trimer chains, respectively.

**B) Magnetic susceptibility χ(*T*) and specific heat at zero-field**

Magnetic susceptibility χ(*T*) as a function of temperature in the range from 2 K to 300 K is shown in Figure 4. The Inverse magnetic susceptibility χ$^{-1}$(*T*) data show a linear behavior at high-*T* above 70 K and then the data steeply fall at low-*T*. Hence, we fitted the data with the below expression in the *T*-range from 70 K to 300 K.

$$\chi = \chi_0 + \frac{C}{T - \theta_{CW}}$$

The temperature independent susceptibility $\chi_0$ was found to be about -1 ×10$^{-4}$ cm$^3$/mol, which is a sum of core diamagnetic susceptibility and Van-Vleck susceptibility. The core diamagnetic susceptibility has been calculated to be -6.3x10$^{-5}$ cm$^3$/mol [21]. The Van-Vleck susceptibility was therefore inferred to be 3.7×10$^{-5}$ cm$^3$/mol, which is in agreement with many cuprate systems. The obtained Curie-Weiss constant is 0.51 cm$^3$ K/mol, which gives an effective magnetic moment of about 2.02 μ$_B$, similar to other Cu-based magnetic materials [22, 23]. The Curie-Weiss temperature ($\theta_{CW}$) is found to be -165 K, indicating the presence of strong AFM correlations in the sample. However, no signature of one dimensional magnetism such as broad maximum in the magnetic susceptibility has been noticed. To further understand the nature of magnetic ground state of this system, we have plotted the χ*T* versus *T* as shown in Figure 4(b). In this plot, the value of χ*T* at 300 K is 0.32 cm$^3$ K/mol, which is smaller than the value expected for free *S* = ½ value 0.3751 cm$^3$ K/mol. This can be attributed to the presence of AFM correlations. This claim is further strengthened from the observation of decreasing χ*T* value with decreasing *T*. At low-*T*, an anomaly is observed at 6 K (= *T*$_c$), indicating the formation of magnetic long-rang order due to the presence of weak inter-layer (3D) couplings.

The specific heat ($C_p$) measurement was carried out in zero-field in the $T$-range from 2 K to 200 K. $C_p(T)$ has two contributions; lattice and magnetic. To understand the magnetic behavior from the specific heat, we have subtracted the lattice contribution using Debye expression mentioned below.

$$C_p = 9rNk_B \sum_{i=1,2} C_i \left(\frac{T}{\theta_D^i}\right)^3 \int_0^{x_D^i} \frac{x^4 e^x}{(e^x - 1)^2} dx$$

Here $r$ is the number of atoms per formula unit, $\theta_D$ is a Debye temperature. The fitting yields $C_1 \approx$ (0.36 ± 0.05), $\theta_{D1} \approx$ (180 ± 5) K, $C_2 \approx$ (0.58 ± 0.05), and $\theta_{D2} \approx$ (650 ± 10) K. The fitted curve was subtracted from the measured specific heat data. As an outcome, the obtained magnetic contribution is plotted in the inset of Figure 5. Similar to $\chi(T)$ data, a sharp anomaly is observed at 6 K. The entropy change $\Delta S_m$ was calculated by integrating the magnetic specific heat divided by $T$ ($C_m/T$) with respect to $T$. The value of $\Delta S_m$ is reached to the maximum value Rln2 at 50 K. The value of $\Delta S_m$ at the transition is found to be 0.4 *J*/mol K, which is only 7 % of total entropy. The observation of remaining large entropy in the paramagnetic region (above $T_c$) suggesting the presence of strong short-range spin correlations. This feature has already been seen in several frustrated spin systems [24, 25].

**C) Magnetic data and specific heat in magnetic fields**

In order to understand the low-$T$ anomaly, we have performed $\chi(T)$ measurements at different magnetic fields. Below 6 K, $\chi(T)$ data at low fields exhibit the splitting between the data measured under the zero-field-cooled (ZFC) and field-cooled (FC) conditions, as shown in inset of Figure 6(a). With increasing $H$, the splitting reduces. The anomaly disappears by the magnetic field of 1 T. The similar behavior is also observed in the $C_p$ data (see Figure 6(b)) under magnetic fields. The magnetic anomaly smears out by the field of 1 T. To further confirm the weak-ferromagnetic nature, we measure the $M(H)$ data at different temperatures. At 2 K, the $M(H)$ plot shows magnetic hysteresis (see inset of Figure 6(b)). The coercive field of 0.05 T and the remanance magnetization of 0.003μ$_B$/mol are observed.

The structure suggests that it has linear chains with Cu-O-Cu bond angles of about 110º to 120º. The super-exchange (SE) couplings are approximately expected to be in the range 130 K – 150 K for this bond angle [26, 27] and the obtained $\theta_{CW}$ is -160 K for this compound. In this picture, one-dimensional magnetism features such as broad maximum is expected at about 100 K in the $\chi(T)$ and 70 K in $C_m(T)$ data [28]. However, we have not seen such features in the magnetic data. Instead, the features of frustrated magnetism is seen with a large frustration parameter $f = \theta_{CW}/T_c \approx 27$ K. The frustrated magnetism could be due to the formation of geometrical network of checkerboard lattice (distorted) by the coupling of chains in the *ab*-plane *via* super-super-exchange (SSE) couplings through O-Te-O atoms. Such SSE interactions were observed to be dominant in many compounds SrCuTe$_2$O$_6$ [25], Na$_2$Cu$_2$TeO$_6$ [29], Cu$_2$A$_2$O$_7$ (A = V, P, and As), *etc* [30-34]. Since the magnetic couplings in the checkerboard layer are highly anisotropic, we might not see the physics of spin-liquid and valance bond solid (VBS) phases, as predicted theoretically [13-17], in this compound.

## IV. CONCLUSION

We have successfully prepared the polycrystalline sample of a new quantum magnet PbCuTeO$_5$. The structure suggests that it has two layers (*ab*-planes) of $S = \frac{1}{2}$ distorted checkerboard lattice, which were formed by the coupling of spin chains. Magnetic susceptibility data exhibit a magnetic anomaly at 6 K despite the presence of large antiferromagnetic Curie-Weiss temperature of about -165 K. Magnetic specific heat data also showed the presence of large magnetic entropy above $T_c$, confirming the existence of short-range spin correlations owing to frustrated magnetic network of the anisotropic checkerboard lattice. Our preliminary magnetic analysis suggests the presence of strong magnetic frustration in this material. The presence of anisotropic and inter-layer couplings might be the reason to observe the magnetic-LRO, instead of the theoretically predicted disordered state for the ideal $S = \frac{1}{2}$ checkerboard or 2D pyrochlore lattice. Further, the NMR and μSR local probe measurements will be helpful to understand the spin dynamics of this quantum magnet.

**Acknowledgments:** B.K. thanks DST INSPIRE faculty award-2014 scheme.

Electronic address: koti.iitb@gmail.com

**FIGURES AND CAPTIONS**

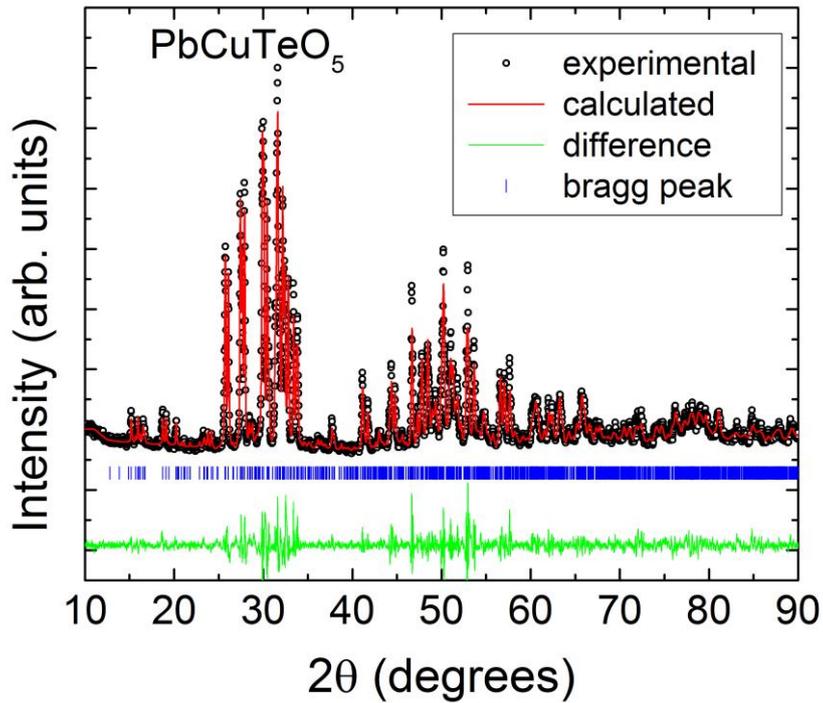

Figure 1: Powder XRD diffraction pattern for the polycrystalline samples of PbCuTeO$_5$. The residual parameters obtained after Rietveld refinement are $R_p \approx 16.4\%$, $R_{wp} \approx 17.9\%$, $R_{exp} \approx 6.80$, and $\chi^2 \approx 6.95$.

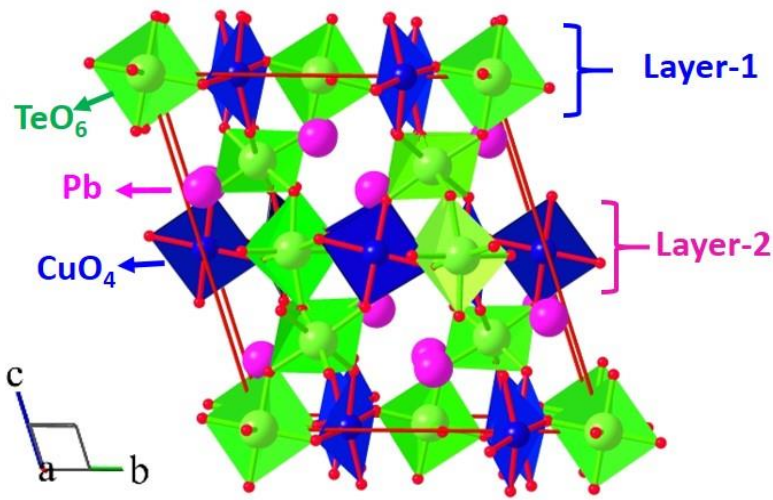

Figure 2: The triclinic unit cell of PbCuTeO$_5$ is built by Pb atoms, TeO$_6$ octahedral and CuO$_4$ squares. It comprises of two kinds of Cu layers; layer-1 ($z=0$) and layer-2 ($z=0.5$).

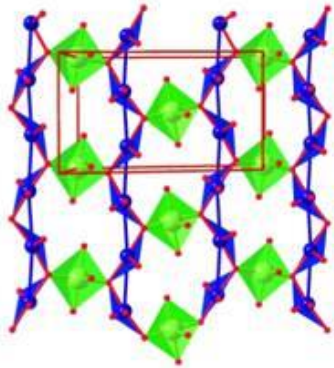
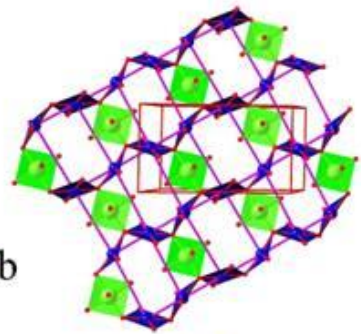
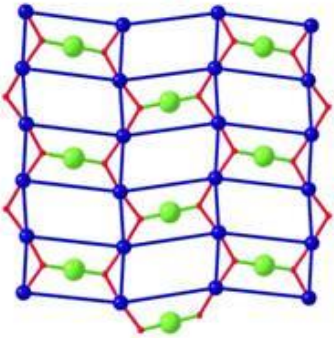
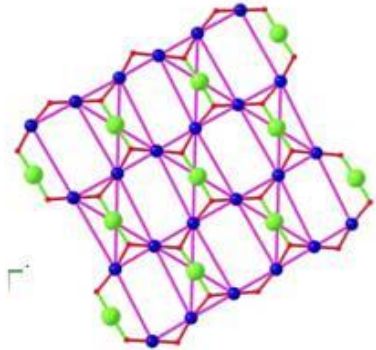
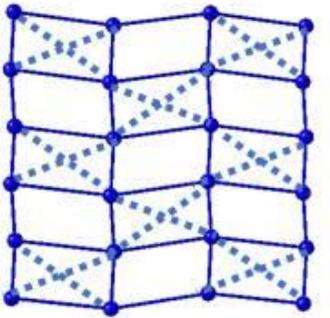
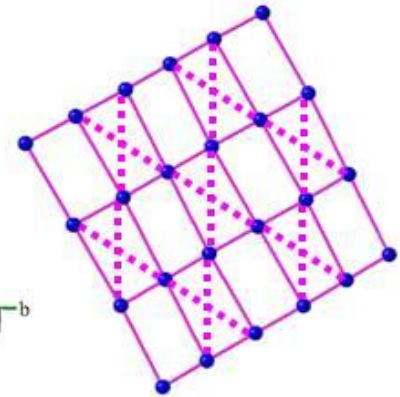

Figure 3: Two Cu layers (*ab*-planes) stacked alternatively. The first layer (z=0) consists of coupled dimer chains, while the second layer (z=0.5) with coupled trimer chains. The Cu-O-Cu chains are coupled each other via O-Te-O bonds forming the checkerboard lattice with anisotropic couplings.

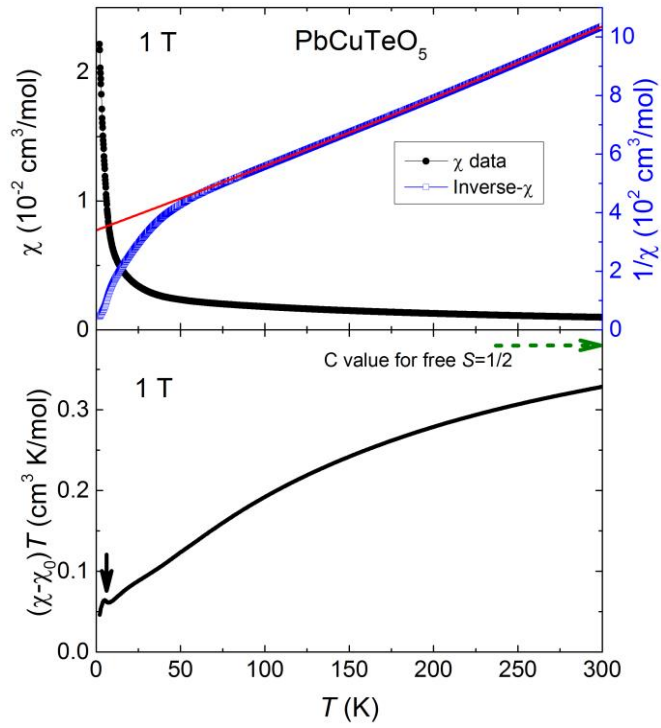

Figure 4: (a) The magnetic susceptibility $\chi(T)$ measured at 1 T and the $\chi^{-1}(T)$ data with a fit to Curie-Weiss law. (b) The plot of $(\chi-\chi_0)T$ versus $T$.

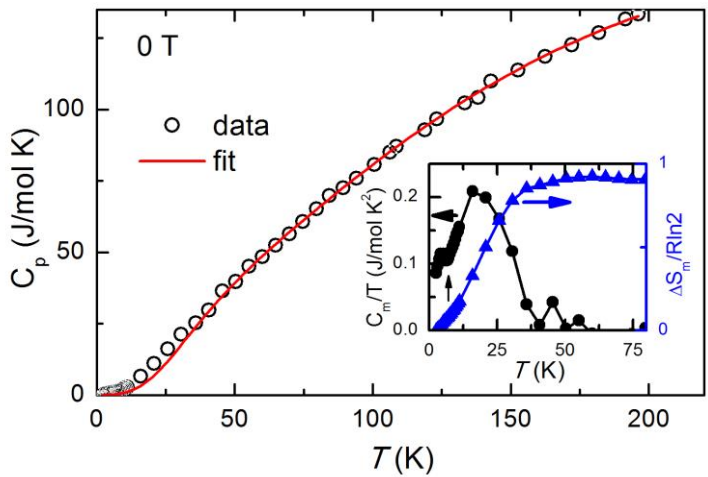

Figure 5: Specific heat ($C_p$) measured in zero-field. The red line indicates the fit to lattice contribution. Inset shows the magnetic specific heat divided by $T$ i.e., $C_m/T$ (left axis) and normalized magnetic entropy ($\Delta S_m/R\ln2$) versus $T$ (right axis).

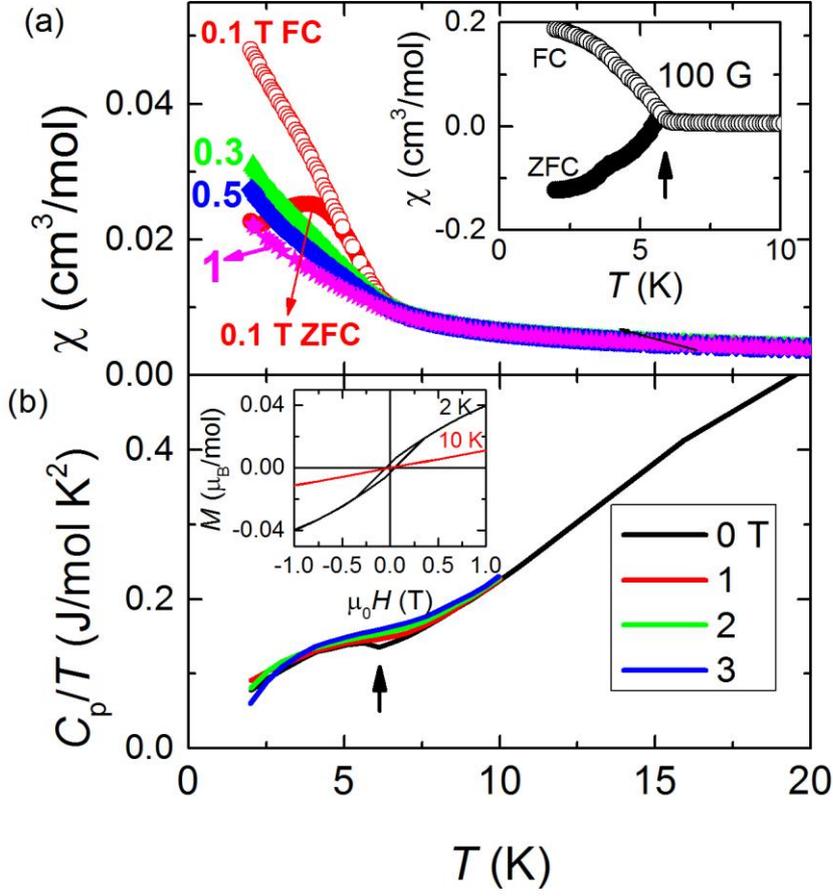

Figure 6: $\chi(T)$ at different fields. The inset shows the plot of ZFC and FC $\chi(T)$ at 100 G. (b) $C_p/T$ versus $T$ at different magnetic fields. Inset shows $M(H)$ data at 2 K and 10 K.

Table 1: The atomic coordinates and the occupancies obtained for PbCuTeO$_5$ after the Rietveld refinement at room temperature using the space group $P$-1 (No. 2).

| Atoms | Wyckoff | x | y | z | Occupancy |
|---|---|---|---|---|---|
| Pb1 | 2i | 0.986 | 0.562 | 0.176 | 1 |
| Pb2 | 2i | 0.490 | 0.448 | 0.329 | 1 |
| Pb3 | 2i | 0.014 | 0.944 | 0.323 | 1 |
| Pb4 | 2i | 0.513 | 0.0602 | 0.181 | 1 |
| Cu1 | 2i | 0.257 | 0.237 | 0.505 | 1 |
| Cu2 | 2i | 0.195 | 0.741 | 0.999 | 1 |
| Cu3 | 1g | 0 | 0.5000 | 0.5000 | 1 |
| Cu4 | 2i | 0.692 | 0.758 | 0.992 | 1 |
| Cu5 | 1f | 0.500 | 0 | 0.500 | 1 |
| Te1 | 1a | 0 | 0 | 0 | 1 |
| Te2 | 2i | 0.235 | 0.743 | 0.502 | 1 |
| Te3 | 2i | 0.467 | 0.764 | 0.248 | 1 |
| Te4 | 1e | 0.500 | 0.500 | 0 | 1 |
| Te5 | 2i | 0.948 | 0.745 | 0.747 | 1 |
| O1 | 2i | 0.981 | 0.650 | 0.461 | 1 |
| O2 | 2i | 0.445 | 0.651 | 0.962 | 1 |
| O3 | 2i | 0.481 | 0.844 | 0.533 | 1 |
| O4 | 2i | 0.019 | 0.919 | 0.840 | 1 |
| O5 | 2i | 0.618 | 0.834 | 0.156 | 1 |
| O6 | 2i | 0.392 | 0.602 | 0.5020 | 1 |
| O7 | 2i | 0.700 | 0.741 | 0.830 | 1 |
| O8 | 2i | 0.077 | 0.889 | 0.501 | 1 |
| O9 | 2i | 0.952 | 0.845 | 0.028 | 1 |
| O10 | 2i | 0.287 | 0.684 | 0.333 | 1 |
| O11 | 2i | 0.201 | 0.769 | 0.668 | 1 |
| O12 | 2i | 0.400 | 0.928 | 0.340 | 1 |
| O13 | 2i | 0.117 | 0.673 | 0.836 | 1 |
| O14 | 2i | 0.202 | 0.470 | 0.012 | 1 |
| O15 | 2i | 0.474 | 0.412 | 0.838 | 1 |
| O16 | 2i | 0.293 | 0.972 | 0.01 | 1 |
| O17 | 2i | 0.127 | 0.421 | 0.352 | 1 |
| O18 | 2i | 0.211 | 0.757 | 0.164 | 1 |
| O19 | 2i | 0.713 | 0.754 | 0.334 | 1 |
| O20 | 2i | 0.217 | 0.177 | 0.342 | 1 |

Table II: The bond angles and the bond lengths of the exchange couplings for PbCuTeO$_5$.

| | Bond path | Bond length (Å) | Bond angle (°) |
|---|---|---|---|
| Layer-1 with dimer chains | Cu4-O9-Cu2 | 3.23 | 115.2 |
| | Cu2-O2-Cu4 | 3.21 | 111.9 |
| | Cu4-O9-Te1-O9-Cu2 | 5.68 | Cu4-O9-Te1= 122.3<br>O9-Te1-O9= 180<br>Te1-O9-Cu2= 112 |
| | Cu4-O9-Te1-O9-Cu4 (diagonal) | 6.70 | Cu4-O9-Te1= 122.3<br>O9-Te1-O9= 180<br>Te1-O9-Cu4= 112 |
| Layer-2 with trimer chains | Cu1-O1-Cu3 | 3.43 | Cu1-O1-Cu3=124.518 |
| | Cu1-O3-Cu-5 | 3.10 | Cu1-O3-Cu5=107.5 |
| | Cu1-O1-Te2-O3-Cu1 | 5.62 | Cu1-O1-Te2=110.4<br>O1-Te2-O3= 175.5<br>Te2-O3-Cu1= 117.5 |
| | Cu1-O1-Te2-O3-Cu5 (diagonal) | 6.49 | Cu1-O1-Te2=110.4<br>O1-Te2-O3= 175.6<br>Te2-O3-Cu5= 112.3 |